\documentstyle[aps,pra]{revtex}
\begin{document}
\draft
\title{Wilson and Kadowaki-Woods Ratios in Heavy Fermions}
\author{Mucio A.Continentino}
\address{Instituto de Fisica,\\
Universidade Federal Fluminense \\
Campus da Praia Vermelha, Niter\'oi, 24.210-340, RJ, Brasil}
\date{\today}
\maketitle

\begin{abstract}
Recently we have shown that a one-parameter scaling, $T_{coh}$, describes
the physical behavior of several heavy fermions in a region of their phase
diagram. In this paper we fully characterize this region, obtaining the
uniform susceptibility, the resistivity and the specific heat in terms of
the coherence temperature $T_{coh}$. This allows for an explicit evaluation
of the Wilson and the Kadowaki-Woods ratios in this regime. These quantities
turn out to be independent of the distance $|\delta|$ to the quantum
critical point (QCP). The theory of the one-parameter scaling corresponds to
a local interacting model. Although spatial correlations are irrelevant in
this case, time fluctuations are critically correlated as a consequence of
the quantum character of the transition.
\end{abstract}




\section{Introduction}

Most of the physical properties of heavy fermions can be attributed to the
proximity of these systems to a quantum critical point (QCP) \cite
{mucio0,hertz}. This zero temperature critical point arises as a competition
between Kondo effect and magnetic order induced by RKKY coupling. In the
magnetically disordered side of the phase diagram a scaling approach reveals
the existence of a new characteristic temperature, the {\em coherence
temperature} $T_{coh}\propto |\delta |^{\nu z}$, below which the system
exhibits Fermi liquid (FL) behavior \cite{mucio1}. In this relation, $%
|\delta |=|J_{Q}-J_{Q}^{c}|$ measures the distance to the $T=0$ critical
point and $\nu $ and $z$ are respectively the correlation length and dynamic
critical exponents. $J_{Q}$ is the coupling between the local moments and $%
J_{Q}^{c}$ its critical value, at which the magnetic instability,
characterized by the wavevector $Q$ occurs. At the QCP, i.e., $|\delta |=0$,
the system does not cross the {\em coherence line} and consequently exhibits 
{\em non-Fermi liquid behavior} down to $T=0$ \cite{mucio0}.

Recently, we have shown that the heavy fermion systems $CeRu_{2}Si_{2}$, $%
CeCu_{6}$, $UPt_{3}$ and $CeAl_{3}$ under pressure (the latter for $P\ge $ $%
1.2$kbars), obey a {\em one-parameter scaling}, i.e., $C/T\propto
T_{coh}^{-1}$, $\chi _{0}(T\rightarrow 0)\propto T_{coh}^{-1}$, $%
A_{R}\propto T_{coh}^{-2}$ and $h_{c}\propto T_{coh}$, where these
quantities are measured in the Fermi liquid regime for $T<<T_{coh}$ \cite
{mucio0,physicab}. The pressure dependence of the coherence temperature, $T_{coh}$,
then determines the pressure variation of the coefficient of the linear term
of the specific heat, of the uniform susceptibility, the coefficient of the $%
T^{2}$ term of the resistivity and of the characteristic pseudo-metamagnetic
field, respectively. In this paper we use the spin-fluctuation theory of a
nearly anti-ferromagnetic metal \cite{moriya} \cite{taki} to fully
characterize this regime and calculate the specific heat, the uniform
susceptibility and the resistivity in the region of the phase diagram where
this type of one-parameter scaling is observed. This allows for an explicit
evaluation of the {\em Wilson ratio} and the {\em Kadowaki-Woods ratio} \cite
{kado,outros} between the coefficient of the $T^{2}$ term in the resistivity
and that of the linear term of the specific heat.

It is important to emphasize that, although the spin fluctuation model that
we use to describe the physical behavior associated with the
antiferromagnetic quantum critical point is a Gaussian theory, this theory
gives the {\em exact description} of the quantum critical behavior \cite
{mucio0}, i.e., the Gaussian exponents associated with the QCP are exact.
The reason is that for the problem considered here, the effective dimension $%
d_{eff}=d+z$. Since the Euclidean dimension $d=3$ and the dynamic exponent $%
z=2$, for antiferromagnetic fluctuations, $d_{eff}>d_{c}=4$, the upper
critical dimension for the magnetic transition. Consequently the Gaussian
fixed point yields the correct zero temperature critical exponents \cite
{millis}.

\section{Specific heat}

We start from the expression for the free energy given by the
spin-fluctuation theory of a nearly antiferromagnetic electronic system \cite
{moriya}\cite{taki}. We use the notation of Ref.\cite{taki}, 
\begin{equation}
f_{sf}=-\frac{3}{\pi }\sum_{\vec{q}}T\int_{0}^{\infty }\frac{d\lambda }{%
e^{\lambda }-1}\tan ^{-1}\left( \frac{\lambda T}{\Gamma _{q}}\right)
\end{equation}
where 
\[
\Gamma _{q}=\Gamma _{L}(1-J_{Q}\chi _{L})+\Gamma _{L}\chi _{L}Aq^{2} 
\]
$\Gamma _{L}$ and $\chi _{L}$ are local parameters defined through the local
dynamical susceptibility, $\chi _{L}(\omega )=\chi _{L}/(1-i\omega /\Gamma
_{L})$\cite{taki} . $J_{Q\text{ }}$, as before, is the $q$-dependent
exchange coupling between $f$-moments and $A$ is the {\em stiffness} of the
lifetime of the spin fluctuations defined by the small wavevector expansion
of the magnetic coupling close to the wavevector $Q$, i.e., $%
J_{Q}-J_{Q+q}=Aq^{2}+\cdots $. Then, $\Gamma _{q}$ can be rewritten as, $%
\Gamma _{q}=\Gamma _{L}\chi _{L}A\xi ^{-2}[1+q^{2}\xi ^{2}]$, where the
correlation length, $\xi =\left( A/|J_{Q}-J_{Q}^{c}|\right) ^{1/2}=\sqrt{%
A/|\delta |}$ diverges at the critical value of the coupling, $%
J_{Q}^{c}=\chi _{L}^{-1}$, with the Gaussian critical exponent, $\nu =1/2$.
Consequently, we have for the specific heat $C/T=-\frac{\partial ^{2}f_{sf}}{%
\partial T^{2}}$, 
\begin{equation}
C/T=\frac{\partial ^{2}}{\partial T^{2}}\{\frac{3}{\pi }\sum_{\vec{q}%
}T\int_{0}^{\infty }\frac{d\lambda }{e^{\lambda }-1}\tan ^{-1}\left( \frac{%
\lambda T\xi ^{z}}{\Gamma _{L}\chi _{L}A(1+q^{2}\xi ^{2})}\right) \}
\end{equation}
where we have identified the dynamic critical exponent, $z=2$, typical of
antiferromagnetic spin fluctuations \cite{moriya}. The argument of the $\tan
^{-1}$ can be written as, $(\lambda T\xi ^{z}/A\Gamma _{L}\chi
_{L})[1-q^{2}\xi ^{2}/(1+q^{2}\xi ^{2})]$. The term in brackets $[...]$ is
always $\le 1$, furthermore the exponential cuts off the contribution for
the integral from large values of $\lambda $, consequently for $(T\xi
^{z}/\Gamma _{L}\chi _{L}A)<<1$, we can expand the $\tan ^{-1}$ for small
values of its arguments. It is easily seen that this condition can be
written as, $T<<T_{coh}$, where the {\em coherence temperature}, 
\[
k_{B}T_{coh}=\Gamma _{L}\chi _{L}|J_{Q}-J_{Q}^{c}|=\Gamma _{L}\chi
_{L}|\delta |^{\nu z} 
\]
in agreement with the scaling theory, since $\nu z=1$ ( $\nu =1/2$ and $z=2$
)\cite{mucio0}. Notice that $T_{coh}$ is {\em independent} of $A$.

We then have for $T<<T_{coh}$,

\begin{equation}
C/T=\frac{\partial ^{2}}{\partial T^{2}}\left[ \frac{3T^{2}}{\pi T_{coh}}%
\int_{0}^{\infty }\frac{d\lambda \lambda }{e^{\lambda }-1}\sum_{\vec{q}}%
\frac{1}{1+q^{2}\xi ^{2}}\right]
\end{equation}
This equation clarifies the physical meaning of the coherence temperature.
It is the characteristic temperature, below which, the free energy is
quadratic in temperature and consequently the system exhibits Fermi liquid
behavior. Changing the $\sum_{\vec{q}}$ into an integral we find ($d=3$), 
\begin{equation}
C/T=\frac{\partial ^{2}}{\partial T^{2}}\left[ \frac{\pi T^{2}\xi ^{(z-d)}}{%
2\Gamma _{L}\chi _{L}A}\frac{4\pi V}{(2\pi )^{3}}\int_{0}^{q_{c}\xi }\frac{%
dyy^{2}}{1+y^{2}}\right]
\end{equation}
which yields 
\begin{equation}
C/T=\frac{\pi \xi ^{(z-d)}}{\Gamma _{L}\chi _{L}A}\frac{4\pi V}{(2\pi )^{3}}%
q_{c}\xi \left( 1-\frac{\tan ^{-1}q_{c}\xi }{q_{c}\xi }\right)
\end{equation}
In the critical regime, $q_{c}\xi >>1$, we obtain the result of Ref.\cite
{taki}, i.e., $C/T=6\pi ^{2}Nk_{B}^{2}/Aq_{c}^{2}$, a non-universal, cut-off
dependent value. In the opposite, sub-critical regime, i.e., for, $q_{c}\xi
<<1$, since $tan^{-1}y\approx y-y^{3}/3+y^{5}/5+\cdots $ for small $y$, we
get \cite{mucio0} 
\begin{equation}
C/T=\frac{\pi \xi ^{(z-d)}}{\Gamma _{L}\chi _{L}A}\frac{4\pi V}{(2\pi )^{3}}%
q_{c}\xi \left[ \frac{1}{3}(q_{c}\xi )^{2}-\frac{1}{5}(q_{c}\xi )^{4}+\cdots
\right]
\end{equation}
The first term is independent of $A$ and yields, 
\begin{equation}
C/T=\frac{\pi Nk_{B}^{2}}{\Gamma _{L}\chi _{L}}\frac{1}{|J_{Q}-J_{Q}^{c}|}=%
\frac{\pi Nk_{B}}{T_{coh}}
\end{equation}
In fact this could have been obtained directly, from Eq.$3$, neglecting the $%
q$-dependence of $\Gamma _{q}$ and with $\sum_{\vec{q}}\rightarrow N$ \cite
{lacroix}. In the equation above the correct units have been restored. In
principle, one may think naively that this limit is not relevant, since the
condition, $q_{c}\xi <<1$, can only be satisfied far away from the quantum
critical point. Note however, that this may be written as, $q_{c}\sqrt{%
A/|\delta |}<<1$, which can be satisfied, {\em either because the system is
far away from the quantum critical point, i.e., }$|\delta |$ {\em is large,
or because }$A${\em \ is small}. Rewriting this condition as $q_{c}/\sqrt{%
|\delta |}<<1/\sqrt{A}$, we notice that, when $A \rightarrow 0$, it holds 
{\em arbitrarily close} to the QCP, i.e., for $\delta $ arbitrarily small.

Table I gives values for the coherence temperature for some heavy fermion
systems, which obey one parameter scaling, obtained from Eq.7 and the
measured specific heat.

\section{Susceptibility and Wilson ratio}

The zero temperature uniform susceptibility of the nearly antiferromagnetic
system in the limit $q_{c}\xi <<1$ can be directly obtained from the
magnetic field $(h)$ dependent, $T=0$, q-independent free energy \cite
{mucio0} \cite{sachdev}, 
\begin{equation}
f_{sf}^{L}=-\frac{3N}{2\pi }\int_{0}^{\omega _{c}}d\omega \tan ^{-1}\left[ 
\frac{\omega +\mu h}{\Gamma _{L}\chi _{L}|J_{Q}-J_{Q}^{c}|}\right]
\end{equation}
Notice that the magnetic field in this expression scales as $(h/h_{c})$
where the characteristic field, $h_{c}\propto T_{coh}$. The uniform
susceptibility $\chi _{0}$ is given by, 
\begin{equation}
\chi _{0}=-\left( \frac{\partial ^{2}f_{sf}^{L}}{\partial h^{2}}\right)
_{h=0}=\frac{3N\mu ^{2}}{2\pi \Gamma _{L}\chi _{L}}\frac{1}{|J_{Q}-J_{Q}^{c}|%
}=\frac{3N\mu ^{2}}{2\pi k_{B}T_{coh}}
\end{equation}
where the limit $\omega _{c}\rightarrow \infty $ has been taken. Note from
the general expression for the uniform susceptibility \cite{taki}, $\chi
^{-1}=\chi _{Q}^{-1}+Aq_{c}^{2}$, where $\chi _{Q}$ is the staggered
susceptibility, that in the limit, $q_{c}\xi <<1$, $\chi =\chi _{Q}=\chi
_{0} $. Then the staggered and uniform susceptibilities coincide in this
limit, being equally enhanced.

The Wilson ratio (WR) is given by 
\[
\frac{\chi _{0}/\mu ^{2}}{C/\pi ^{2}k_{B}^{2}T}=\frac{3}{2}=1.5 
\]
which turns out to be a universal number since the dependence on the
distance to the critical point, $|J_{Q}-J_{Q}^{c}|$ and on the dimensionless
quantity $\Gamma _{L}\chi _{L}$ cancels out. We emphasize that the above
result for $\chi _{0}$ is valid in the regime, $q_{c}\xi <<1$, that is, if
the system satisfies the condition, $q_{c}/\sqrt{|J_{Q}-J_{Q}^{c}|}<<1/\sqrt{%
A}$. The experimental value for the WR in $CeRu_{2}Si_{2}$, given in Table
I, is in fair agreement with the theory. The same is not true for the other
systems. Since there is a reasonable spread on the experimental data for
different samples of these materials, at this point, the best indication
that they are in the sub-critical regime is the fact that they obey
one-parameter scaling.

\section{Resistivity and Kadowaki-Woods ratio}

The resistivity due to spin fluctuations in the regime $q_{c}\xi <<1$ is
given by \cite{mucio0} \cite{lederer} 
\begin{equation}
\rho =\rho _{0}\frac{1}{T}\int_{0}^{\infty }d\omega \frac{\omega \Im m\chi
_{Q}(\omega )}{(e^{\beta \omega }-1)(1-e^{-\beta \omega })}
\end{equation}
where 
\begin{equation}
\Im m\chi _{Q}(\omega )=\chi _{Q}^{s}\frac{\omega \xi _{L}^{z}}{1+(\omega
\xi _{L}^{z})^{2}}
\end{equation}
with 
\[
\chi _{Q}^{s}= \frac{1}{|J_{Q}-J_{Q}^{c}|} 
\]
and 
\[
\xi _{L}^{z}=\frac{\chi _{Q}^{s}}{\Gamma _{L}\chi _{L}} 
\]
The quantity $\rho _{0}$ is given by, 
\[
\rho _{0}=(\frac{J}{W})^{2}\frac{m_{c}}{N_{c}e^{2}\tau _{Fc}}(N/N_{c}) 
\]
where $J$ is the coupling constant per unit cell between localized and
conduction electrons. $W$, $m_{c}$ and $N_{c}$ are the bandwidth, the mass
and the number of conduction electrons per unit volume with Fermi momentum $%
k_{Fc}$, such that, $\hbar {\tau _{Fc}}^{-1}=\hbar ^{2}k_{Fc}^{2}/2m_{c}$
and $N$ is the number of atoms per unit volume.

Using the definitions above, we can rewrite the spin- fluctuation
resistivity as, $\rho =\rho _{0}\Gamma _{L}\chi _{L}R(\tilde{T})$ , where 
\begin{equation}
R(\tilde{T})=\frac{1}{\tilde{T}}\int_{0}^{\infty }d\tilde{\omega}\frac{1}{%
(e^{\tilde{\omega}/\tilde{T}}-1)(1-e^{-\tilde{\omega}/\tilde{T}})}\frac{%
\tilde{\omega}^{2}}{1+\tilde{\omega}^{2}}
\end{equation}
with $\tilde{\omega}=\omega \xi _{L}^{z}$ and $\tilde{T}=T\xi _{L}^{z}$.
For, $T<<T_{coh}=(\Gamma _{L}\chi _{L}/k_{B})|J_{Q}-J_{Q}^{c}|$, we have, $%
R(T<<T_{coh})\approx \frac{\pi ^{2}}{3}(\frac{T}{T_{coh}})^{2}$ and finally, 
\begin{equation}
\rho (T<<T_{coh})=\rho _{0}\Gamma _{L}\chi _{L}\frac{\pi ^{2}}{3}(\frac{T}{%
T_{coh}})^{2}=A_{R}T^{2}
\end{equation}
where 
\[
A_{R}=\frac{\rho _{0}\pi ^{2}}{3}\frac{k_{B}^{2}}{\Gamma _{L}\chi _{L}}\frac{%
1}{|J_{Q}-J_{Q}^{c}|^{2}} 
\]
Notice that, {\em the same coherence temperature}, $T_{coh}$, appears in the
transport properties. In this case, it is the characteristic temperature
below which the resistivity varies quadratically with temperature, as
appropriate to a FL and that sets the scale for this contribution. Note that
there are two characteristic time scales in the problem considered here,
namely $\xi ^{z\text{ }}$and $\xi _{L}^{z}$. While the former vanishes in
the local limit the latter may still diverge.

The Kadowaki-Woods ratio \cite{kado}, $A_{R}/({C/T})^{2}$ is given by, 
\begin{equation}
\frac{A_{R}}{(C/T)^{2}}=\frac{\rho _{0}\Gamma _{L}\chi _{L}}{3(Nk_{B})^{2}}
\end{equation}
which depends on the local parameters, $\Gamma _{L}\chi _{L}$, consequently
on the nature of the magnetic ion ($4f$, $5f$ or $d$, for example), but{\em %
\ not} on the distance to the critical point, $|J_{Q}-J_{Q}^{c}|$. From the
equation above and the expression for $\rho _{0}$ we obtain, $(\Gamma
_{L}\chi _{L})\left[ \frac{J}{nW}\right] ^{2}=\frac{6.55\times 10^{-3}}{%
n^{2/3}v_{0}^{1/3}}\frac{A_{R}}{(C/T)^{2}}$, with, $n=(N_{c}/N)$, the
average number of conduction electrons per atom and $v_{0}^{1/3}$, the
average atomic radius given in $cm$. The values of $(\Gamma _{L}\chi
_{L})(J/W)^{2}$ obtained from the experimental results and with $n=1$, are
given in Table I. If we use $(\Gamma _{L}\chi _{L})=1/2\pi $, the value for $%
S=1/2$ \cite{taki}, we get, $(J/W)=1.6$, $1.8$ and $2.1$ for $CeRu_{2}Si_{2}$%
, $UPt_{3}$ and $CeCu_{6}$, respectively, which are too large. On the other
hand, extending the Korringa relation \cite{taki} for arbitrary spin, i.e., $%
\Gamma _{L}\chi _{L}=2(\mu /g_{J}\mu _{B})^{2}/3\pi $, where $\mu $ is the
experimental magnetic moment (in $\mu _{B}$) given in Table I, we get the
values for $(J/W)$ also shown in this Table ($n=1$). These values are in
agreement with those expected for non-magnetic heavy fermions \cite{coqblin}%
, although the value of $(J/W)_{c}$ separating the magnetic from the Fermi
liquid ground states in heavy fermions is still unknown. Takimoto and Moriya 
\cite{taki2} have also calculated the Kadowaki-Woods ratio for heavy
fermions and obtained that, not too close to the QCP, it is nearly
independent of $|\delta |$.

Notice that for $T>>T_{coh}$, $R(T)\approx \frac{\pi T}{2T_{coh}}$ and the
resistivity varies linearly with temperature. It is given by, 
\[
\rho (T>>T_{coh})=\rho _{0}\Gamma _{L}\chi _{L}\frac{\pi }{2}\frac{T}{T_{coh}%
} 
\]

We point out that in the $q$-dependent, critical regime, i.e., for $q_{c}\xi
\gg 1$, also $\rho =A_{R}^{M}T^{2}$, at low temperatures, but the
coefficient $A_{R}^{M}\propto |J_{Q}-J_{Q}^{c}|^{-1/2}$ \cite{taki} and
consequently does not scale as $T_{coh}^{-2}$, in disagreement with the
experimental results in the heavy fermion systems investigated here \cite
{mucio0}. Also in the critical regime, the thermal mass, $C/T=6\pi
^{2}Nk_{B}^{2}/Aq_{c}^{2}$ \cite{taki}, as shown before, and the uniform
susceptibility, $\chi =N\mu ^{2}/Aq_{c}^{2}$ \cite{taki} both saturate at
non-universal, cut-off dependent values \cite{taki}, and do not scale as
observed experimentally. In this critical regime these quantities are
controlled by the {\em stiffness} $A$. On the other hand, in the local
limit, $q_{c}\xi <<1$, the relevant Fermi liquid parameters are universal
since they are independent of the cut-off $q_{c}$ and the stiffness $A$.
They are determined by the coherence temperature (Eqs.$7$, $9$ and $12$),
essentially by the distance of the system to the QCP.

It should be clear by now that the spin fluctuation theory of nearly
antiferromagnetic systems, in the sub-critical regime $q_{c}\xi \ll 1$,
gives rise to the one-parameter scaling observed in the heavy fermions
discussed here. We have obtained, $C/T\propto T_{coh}^{-1}$, $\chi
_{0}\propto T_{coh}^{-1}$, $A_{R}\propto T_{coh}^{-2}$ and $h_{c}\propto
T_{coh}$, as found experimentally in the pressure experiments. The heavy
fermions, $CeRu_{2}Si_{2}$, $CeCu_{6}$, $UPt_{3}$ and $CeAl_{3}$ ($P>1.2$%
kbars) are then close, {\em but not too close}, to the QCP such that they
satisfy the condition $q_{c}\xi \ll 1$ or $q_{c}/\sqrt{|\delta |}<<1/\sqrt{A}
$. In the phase diagram of Fig.1, they are close but to the right of the
line $(q_{c}\xi )^{-1}=1$, shown in this figure. Note that pressure used in
the experiments drives these systems further away to the right of this line
where the results obtained here are valid.

\section{Conclusion}

We have shown that the linear term of the specific heat, the uniform
susceptibility and the coefficient of the $T^{2}$ term of the resistivity of
a nearly antiferromagnetic metallic system in the Fermi liquid regime, for $%
q_{c}\xi \ll 1$, can be written in terms of the coherence temperature. The
condition, $q_{c}\xi \ll 1$, or $q_{c}\left( A/|\delta |\right) ^{1/2}\ll 1,$
does not imply that the system is far away from the critical point since it
can be satisfied for arbitrarily small $|\delta |=|J_{Q}-J_{Q}^{c}|$, for $A$
sufficiently small.

The spin fluctuation theory in the sub-critical regime $q_{c}\xi \ll 1$
corresponds to a {\em local interacting model} as the correlation length may
become smaller than the distance among spins. Formally, this regime is
equivalent to a problem in Euclidean dimension $d=0$, consistent with the
local character of the model in this limit, but with effective
dimensionality $d_{eff}=z=2$ \cite{physicab}. In the local model the
exponent $\alpha $, defined through $f_{sf}^{L}\propto |\delta |^{2-\alpha }$
for $\delta \rightarrow 0$, takes the value $\alpha =1$ as can be seen from
Eq.8 with $h=0$. From the divergence of the characteristic time $\tau
_{L}\propto \xi _{L}^{z}\propto |\delta |^{-1}$ in Eq.11, which yields $\nu
z=1$, we get that the quantum hyperscaling relation \cite{mucio1} $2-\alpha
=\nu (d+z)$ is indeed satisfied for $d=0$. In spite that spatial
correlations are irrelevant in this regime, time fluctuations are critically
correlated due to the quantum character of the transition.

In the regime $q_{c}\xi <1$, the system obeys a one-parameter scaling, i.e., 
$C/T\propto T_{coh}^{-1}$, $\chi _{0}\propto T_{coh}^{-1}$, $A_{R}\propto
T_{coh}^{-2}$, $h_{c}=T_{coh}$, with $T_{coh}\propto \xi ^{-z}$, as we have
shown. We can conclude that the heavy fermion systems $CeRu_{2}Si_{2}$, $%
CeCu_{6}$ and $UPt_{3}$, for $P\geq 0$ and $CeAl_{3}$, for $P>1.2$kbars,
belong to a region of Doniach `s phase diagram where the condition, $%
q_{c}\xi <1$, is satisfied as evidenced by the one-scaling parameter
observed in these materials \cite{mucio0}\cite{nota}. These systems are to
the right of the line $(q_{c}\xi )^{-1}=1$, in the phase diagram of Fig.1.

As the quantum critical point is further approached, $q_{c}\xi \rightarrow
\infty $ and the full $q$-dependence of the dynamic susceptibility must be
taken into account. This critical regime has been described by Takimoto and
Moriya \cite{taki}, but it clearly does not yield the one-parameter scaling
obtained in the pressure experiments on the systems above.

Eventually for $|\delta |=0$, the coherence temperature vanishes and the FL
regime is never reached. The thermodynamic properties in this {\em non-Fermi
liquid} regime have been obtained by Moriya \cite{taki}, Millis \cite{millis}
and in Ref. \cite{mucio0}, for the case the antiferromagnetic transitions at 
$T>0$ are described by non-Gaussian exponents. In fact, although the
spin-fluctuation theory gives the exact exponents of the quantum critical
point, this is not the case for the finite temperature antiferromagnetic
transitions which occur in the region of the Kondo lattice phase diagram for 
$(J/W)<(J/W)_{c}$, or $J_{Q}<J_{Q}^{c}$.

\acknowledgements I would like to thank Conselho Nacional de Desenvolvimento
Cientifico e Tecnologico for partial financial support.

\begin{table}[h]
\begin{center}
\par
\begin{tabular}{|c|c|c|c|}
\hline
& $CeRu_2Si_2$ & $CeCu_6$ & $UPt_3$ \\ \hline\hline
$\mu(\mu_B)$ & $2.5$ & $2.5$ & $3.0 $ \\ \hline
$v_0^{1/3}(\times 10^{-8} cm) $ & $4.4$ & $4.7$ & $4.1$ \\ \hline
$T_{coh}(K)^{*}$ & $67$ & $15$ & $58$ \\ \hline
$\chi/\mu^2$ $(\times 10^{35} erg^{-1}cm^{-3})$ & $5.8$ & $7.9$ & $2.5$ \\ 
\hline
$\frac{C/T}{\pi^2 k_B^2}$ $(\times 10^{35} erg^{-1} cm^{-3})$ & $3.9$ & $%
13.0 $ & $5.6$ \\ \hline
$\frac{\chi/\mu^2}{C/T \pi^2 k_B^2}$ & $1.46$ & $0.59$ & $0.44$ \\ \hline
$A_R$ $(\mu \Omega cm K^{-2})$ & $0.40$ & $14.4$ & $0.7^{**}$ - $1.6^{***}$
\\ \hline
$\frac{A_R}{(C/T)^2}$ $(\times 10^{-5} \mu \Omega cm (\frac{mol K}{mJ})^2)$
& $0.27$ & $0.51$ & $0.34^{**}$ - $0.79^{***}$ \\ \hline
$(\Gamma_L \chi_L)(J/W)^2$ & $0.40$ & $0.71$ & $0.54^{**}$ - $1.25^{***}$ \\ 
\hline
$(J/W)$ & $0.47$ & $0.63$ & $0.39^{**}$ - $0.59^{***}$ \\ \hline
\end{tabular}
\par
\end{center}
\caption{Parameters for some heavy fermion systems. All experimental data
are taken from \protect\cite{thompsom} and references therein. $(*)$
obtained from Eq.7 for the specific heat. $(**)$ along the c-axis. $(***)$
in the basal plane. For the transport parameter, the average number of
conduction electrons per atom, $n=1$.}
\end{table}

{\bf Figure Caption}

Phase diagram of heavy fermions. The quantum critical point is located at $%
(1/q_c \xi)=0$. Below the coherence line, $T_{coh}$ the system behaves as a
Fermi liquid. The line $(1/q_c \xi)=1$ separates the true critical regime, $%
(1/q_c \xi) \ll 1$, from the local regime where one-parameter scaling is
observed. The crossover along this line is smooth and can be seen as a
dimensional crossover from $d=0$, the local regime, to $d=3$ (see text). The
materials investigated here are to the right of the line $(1/q_c \xi)=1$


\begin{references}
\bibitem{nota}  We do not calculate numerically this condition since $q_{c}$
and $A$ are unknown. It can in fact be relaxed, as the local model already
holds for $q_{c}\xi \lesssim 1$ \cite{mucio0}.
\end{references}
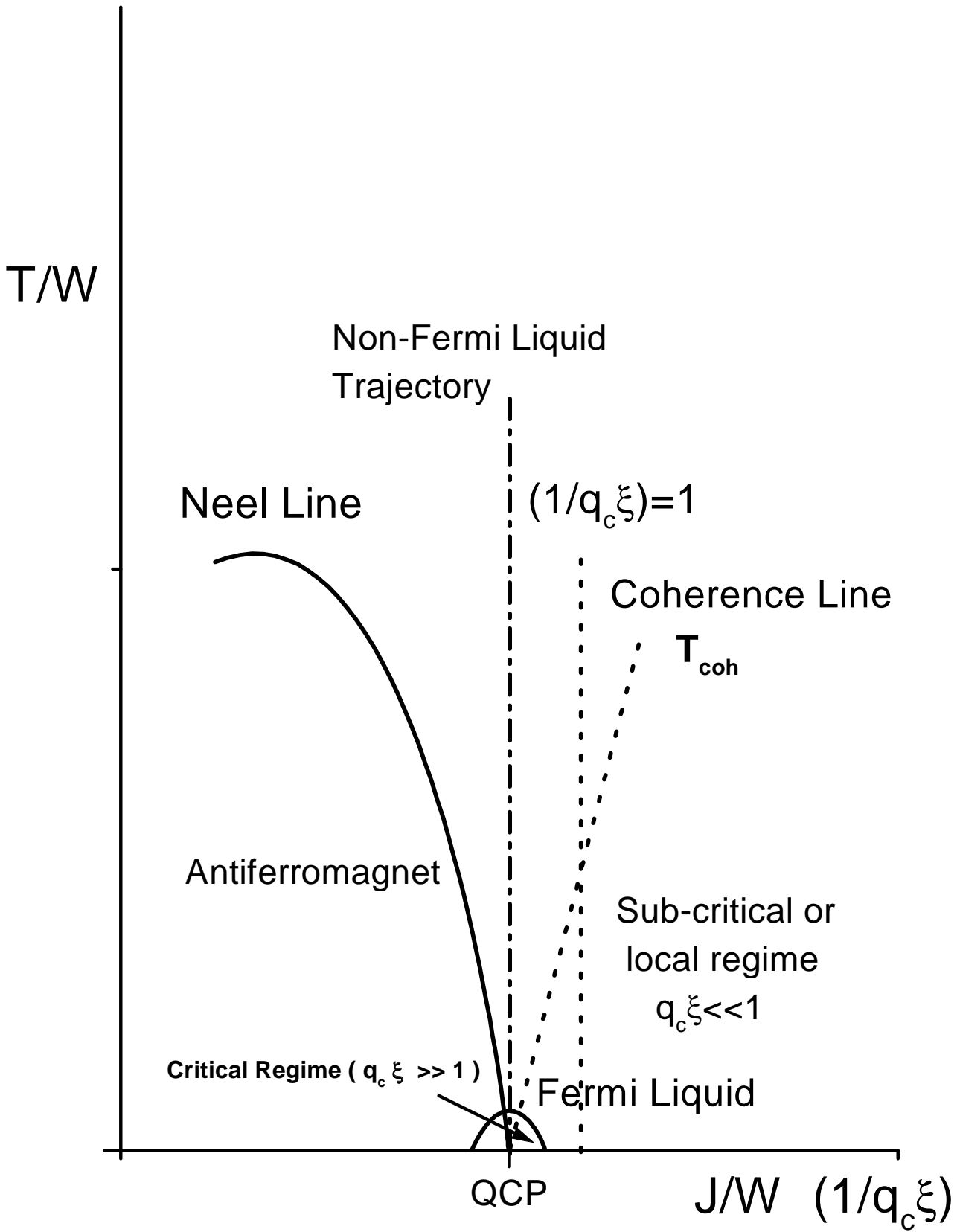
\end{document}